\documentclass[%
 reprint,
superscriptaddress,
 amsmath,amssymb,
 aps,
]{revtex4-2}

\usepackage{graphicx}
\usepackage{dcolumn}
\usepackage{bm}

\usepackage{comment}

\begin{document}

\preprint{APS/123-QED}

\title{Squeeze Free Space with Nonlocal Flat Optics}

\author{Cheng Guo}
 \affiliation{Department of Applied Physics, Stanford University, Stanford, California 94305, USA}

\author{Haiwen Wang}
 \affiliation{Department of Applied Physics, Stanford University, Stanford, California 94305, USA}
 
\author{Shanhui Fan}
\email{shanhui@stanford.edu}
\affiliation{%
 Ginzton Laboratory and Department of Electrical Engineering, Stanford University, Stanford, California 94305, USA
}

\date{\today}

\begin{abstract}
There has been substantial interest in miniaturizing optical systems by flat optics. However, one essential optical component, free space, fundamentally cannot be substituted with conventional \emph{local} flat optics with  \emph{space}-dependent transfer functions, since the transfer function of free space is  \emph{momentum}-dependent instead. Overcoming this difficulty is important to achieve utmost miniaturization of optical systems. In this work, we show that free space can be substituted with \emph{nonlocal} flat optics operating directly in the \emph{momentum} domain. We derive the general criteria for an optical device to replace free space, and provide a concrete design of a photonic crystal slab device. Such a device can substitute much thicker free space with a compression ratio as high as $144$. Our work paves the way for utmost miniaturization of optical systems using  combination of local and nonlocal flat optics.  

\end{abstract}

\maketitle


\section{\label{sec:Intro}Introduction}

Recently, there has been significant progress in flat optics aiming to miniaturize optical systems by replacing conventional optical components~\cite{yu2014,arbabi2015,khorasaninejad2016a,chen2016}. Such flat optics typically achieves wavefront shaping through \emph{local} control of phase, amplitude and polarization in the \emph{spatial} domain. This approach enables compact devices including metalenses, beam deflectors and holograms, with potential applications such as flat displays, wearable optics,  and lightweight imaging systems~\cite{genevet2017}. 

However, there is one essential optical component that has been long-overlooked in flat optics: free space. Free space is an essential part of most optical systems, and usually constitutes a great portion of the system volume. The utmost miniaturization of optical systems therefore requires significant reduction of  free space: a truly miniaturized imaging system needs not only compact flat lenses, but also squeezed free space.

Free space fundamentally can not be substituted with conventional \emph{local} flat optics  characterized by \emph{space}-dependent transfer functions, since the free space propagation has a \emph{momentum}-dependent transfer function  instead.  On the other hand, recently there has been significant interest in lensless Fourier optics using \emph{nonlocal}  photonic nanostructures with tranfer functions in the \emph{momentum} domain. This approach enables important functionalities in optical analog processing including optical differentiation and filtering using compact devices ~\cite{Silva2014,pors2015analog,youssefi2016analog,zhu2017plasmonic,Guo2018,guo2018isotropic,zhu2019generalized,davis2019metasurfaces,zhou2019optical,wang2020compact}. 

In this work, we show that free space can be substituted with \emph{nonlocal} flat optics. We derive the general criteria for a device to replace free space, and provide a concrete design of such a device utilizing photonic Fano resonances. The device can substitute much thicker free space with a compression ratio as high as $144$. It can be placed anywhere in the optical path to reduce free space as needed. Our work therefore opens up significant opportunities in miniaturizing optical systems by squeezing free space with nonlocal flat optics.

The rest of this paper is organized as follows. In Sec.~\ref{sec:theory} we provide a theoretical analysis of replacing free space with flat optics  using Fano resonaces. In Sec.~\ref{sec:numerical} we provide numerical demonstration of a concrete photonic crystal slab design to realize such a functionality. We conclude in  Sec.~\ref{sec:conclusion}.

\begin{figure}[htbp]
    \centering
    \includegraphics[width=1.0\columnwidth]{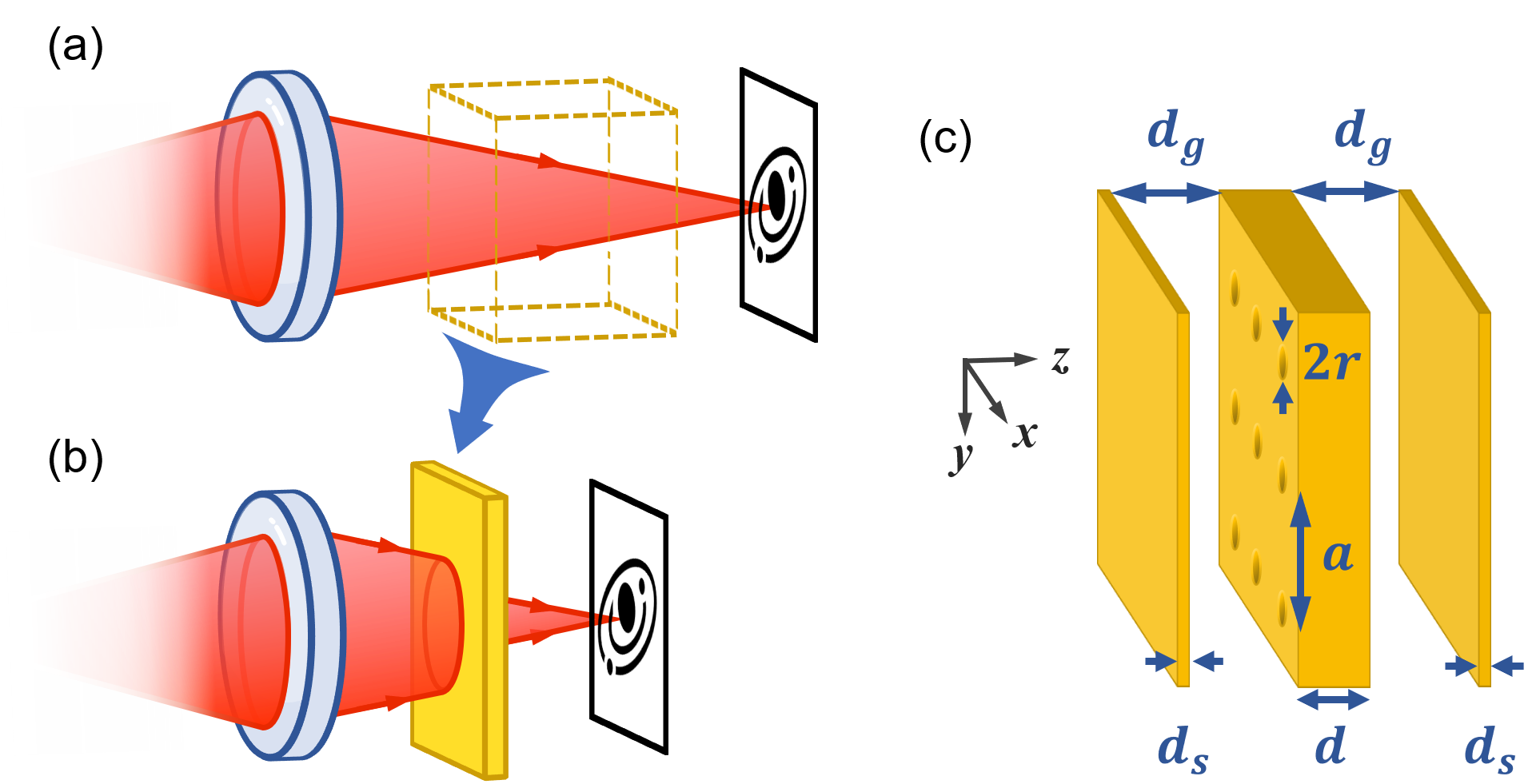}
    \caption{(a) A general optical system, showing the lens (blue), light beam (red) and a formed image. The free space propagation is highlighted as the dashed box. (b) The free space propagation can be replaced by a nonlocal optical device, which has a significantly reduced physical length, but performs the same functionality as the free space. (c) Properly designed photonic crystal slab can serve as a substitute for free space propagation. The geometric parameters are $r=0.111a$, $d=0.55a$, $d_s=0.07a$, $d_g=0.94a$. $a$ is the periodicity along both $x$ and $y$ directions. The yellow regions correspond to material with permittivity $\varepsilon=12$. }
    \label{fig:scheme}
\end{figure}

\section{\label{sec:theory}Theoretical analysis}

In this section, we provide a theoretical analysis on substituting free space with flat optics.

First we briefly examine the propagation of light in free space. Consider a monochromatic optical wave of wavelength $\lambda$ and complex amplitude $U(x,y,z)$ in the free space between the input plane $z=0$ and the output plane $z=d$.  The free space propagation that maps $U(x,y,0)$ to $U(x,y,d)$ is a shift-invariant linear system, since the Helmholtz equation that governs $U(x,y,z)$ is linear,  and free space is invariant under spatial translation in the $xy$-plane. Such a shift-invariant linear system is characterized by its transfer function~\cite{saleh2007}:
\begin{equation}
    \label{eq:transfer_function}
    H(k_x,k_y)=\exp(-i d\sqrt{k_0^2-k_x^2-k_y^2}),
\end{equation}
where $\bm{k}=(k_x,k_y)$ is the in-plane wavevector, and $k_0 = 2\pi/\lambda$ is the angular wavenumber in free space. $H(k_x,k_y)$ is a circularly symmetric complex function of  $k_x$ and $k_y$. A plane wave component with $k_x^2+k_y^2\leq k_0^{2}$ is propagating. For such a wave, the magnitude $|H(k_x,k_y)|=1$ and the phase $\operatorname{arg}\{H(k_x,k_y)\}$ is wavevector dependent. A plane wave component with    $k_x^2+k_y^2>k_0^2$ is evanescent. For such a wave,  $H(k_x,k_y)$ is  exponentially decaying. 

Given the input field $U(x,y,0)$, the output field $U(x,y,d)$ is determined as: 
\begin{equation}
    \label{eq:prop}
\begin{split}
    U(x,y,d) = \iint\limits _{-\infty} ^{+\infty} &H(k_x,k_y)\tilde U(k_x,k_y,0)\\    &\exp[-i(k_x x+k_y y)] \, \mathrm{d} k_x \mathrm{d} k_y,
    \end{split}
\end{equation}
where 
\begin{equation}
    \label{eq:ft}
    \tilde U(k_x,k_y,0) = \frac{1}{(2\pi)^2}\iint\limits _{-\infty} ^{+\infty} U(x,y,0)
    \exp[i(k_x x+k_y y)] \, \mathrm{d} x \mathrm{d} y.
\end{equation}

In many cases of interest, the optical waves are paraxial where the input field  $U(x,y,0)$ contains only wavevector components for which  $k_x^2+k_y^2\ll k_0^2$. The transfer function in Eq.~(\ref{eq:transfer_function}) then can be simplified as
\begin{equation}
    \label{eq:fresnel_approx}
    H(k_x,k_y)\approx H_0\exp[i\frac{\lambda d}{4\pi}(k_x^2+k_y^2)],
\end{equation}
where $H_0 = \exp(-ik_0 d)$ is a global phase. This is the well known Fresnel approximation~\cite{goodman2005}. This approximation is applicable if 
\begin{equation}
    \label{eq:fresnel_condition}
    \frac{(k_x^2+k_y^2)^2 \lambda^3 d}{64\pi^4}\ll 1.
\end{equation}

In order to squeeze free space of a propagation distance $d$, our objective will be to create an optical device with the same transfer function of  Eq.~(\ref{eq:fresnel_approx}), but with a physical thickness that is much lesser than $d$. Since the transfer function of Eq.~(\ref{eq:fresnel_approx}) describes a wavevector-dependent phase shift, with no mixing between different wavevector components, the required optical device must be periodic in order to preserve the wavevector as light transmits through the device. This is in contrast with standard metasurfaces, which are characterized by a space-dependent phase shift, and hence does not preserve the wavevector in the transmission process~\cite{kwon2018,guo2018isotropic}. In addition, to achieve Eq.~(\ref{eq:fresnel_approx}) ideally we will need an all-pass filter with unity amplitude transmission coefficient for all wavevector components.


In this work, we show show that the transfer function of Eq.~(\ref{eq:fresnel_approx}) can be achieved utilizing the phase response of Fano resonances~\cite{zhou2014,limonov2017,zhou2019a}. We consider a single band of guided  resonances in a 2D photonic crystal slab. Assuming that the slab has mirror symmetry in the vertical direction ($z$ direction), the transmitted and reflected amplitudes near resonant frequencies can be expressed as \cite{Fan2002}:
\begin{align}
    \label{eq:fano}
    t(\omega,\bm{k}) &= t_d + f\frac{\gamma(\bm{k})}{i(\omega-\omega(\bm{k}))+\gamma(\bm{k})},\\
    r(\omega,\bm{k}) & = r_d \pm f\frac{\gamma(\bm{k})}{i(\omega-\omega(\bm{k}))+\gamma(\bm{k})}\label{eq:fano-r}
\end{align}
where $t_d$ and $r_d$ are the direct transmission and reflection coefficients,  $\omega(\bm{k})$ and $\gamma(\bm{k})$ are the center frequencies and radiative linewidths of the guided resonance band, and $f$ is the normalized complex amplitude. The plus/minus sign in Eq.~\ref{eq:fano-r} corresponds to even/odd mode with respect to $z-$mirror plane, respectively.

For such a system, energy conservation requires \cite{Fan2002,zhao2019d}  
\begin{equation}
    \label{eq:f_t_d}
    f = -(t_d\pm r_d),
\end{equation}
\begin{equation}
\label{eq:rd2_td2}
    |r_d|^2 + |t_d|^2 = 1,
\end{equation}
\begin{equation}
    \label{eq:rd_td}
    r_d/t_d \equiv -iq, \ q\in \mathbb{R}.
\end{equation}
Denote dimensionless frequency detuning as
\begin{equation}
    \label{eq:detuning}
    \Omega(\omega,\bm{k}) = \frac{\omega-\omega(\bm{k})}{\gamma(\bm{k})},
\end{equation}
then Eq.~(\ref{eq:fano}) with Eqs.~(\ref{eq:f_t_d}-\ref{eq:rd_td}) becomes 
\begin{equation}
    \label{eq:Fano_Omega}
    t(\Omega) = t_d \frac{i(\Omega\pm q)}{1+i\Omega}
\end{equation}
which is a typical Fano lineshape function with $\pm q$ being the asymmetric parameter \cite{Miroshnichenko2010}, and $|t_d|=1/\sqrt{1+q^2}$. Near the resonance, the transmission magnitude  varies between $|t|=0$ at $\Omega=\mp q$ and $|t|=1$ at $\Omega=\pm 1/q$. The phase also exhibits a rapid variation as a function of frequency:
\begin{equation}
    \label{eq:t_arg}
    \operatorname{arg}[t(\Omega)] = \operatorname{arg}(t_d) + \frac{\pi}{2} \operatorname{sgn}(\Omega\pm q)-\arctan{(\Omega)},
\end{equation}
where the first term corresponds to the phase of the directly transmitted amplitude, which is slowly varying in $\omega$ and $\bm{k}$; the second term  describes the abrupt $\pi$ phase jump at the zero transmission frequency; the last term  corresponds to the rapid variation of the phase due to the guided resonance.

Now, we show the band dispersion of guided resonances can be used to realize the quadratic phase response as described by Eq.~(\ref{eq:fresnel_approx}). We consider a specific band of guided resonances near $\bm{k}=\bm{0}$ with isotropic band dispersion
\begin{equation}
    \label{eq:dispersion}
    \omega(\bm{k})\approx\omega_0+\alpha(k_x^2+k_y^2),
\end{equation}
and small dispersion of linewidths
\begin{equation}
    \label{eq:gamma}
    \gamma(\bm{k})\approx \gamma_0.
\end{equation}
Then
\begin{equation}
    \label{eq:Omega_expansion}
    \Omega(\omega,\bm{k}) \approx \Omega_0- \frac{\alpha}{\gamma_0}(k_x^2+k_y^2),
\end{equation}
where 
\begin{equation}
    \label{eq:Omega0}
    \Omega_0 \equiv \Omega(\omega,\bm{0}) = \frac{\omega - \omega_0}{\gamma_0}.
\end{equation}
Assuming $\Omega(\omega,\bm{k})\neq \mp q$ for all the relevant $\bm{k}$,
\begin{equation}
    \label{eq:arg_expansion}
    \operatorname{arg}[t(\Omega)] \approx \operatorname{arg}[t(\Omega_0)] + \frac{\alpha}{\gamma_0 (1+\Omega_0^2)}(k_x^2+k_y^2),
\end{equation}
which gives the the same quadratic dependent of phase on wavevector, as required in Eq.~(\ref{eq:fresnel_approx}).

Finally, we note that by choosing suitable $q$ and operating frequency $\omega\approx\pm 1/q$ correspondingly, in general one can  achieve near-unity transmission  $|t(\Omega(\omega,\bm{k}))|\approx1$ for all the relevant $\bm{k}$. Therefore, the  transfer function for  monochromatic waves with frequency $\omega$ becomes
\begin{equation}
    \label{eq:t_kx_ky}
    t(k_x,k_y) = t(\Omega_0) \exp[i\frac{\alpha}{\gamma_0 (1+\Omega_0^2)}(k_x^2+k_y^2)]
\end{equation}
where $t(\Omega_0)$ is a global phase with $|t(\Omega_0)|=1$. Eq.~(\ref{eq:t_kx_ky}) has exactly the same form as Eq.~(\ref{eq:fresnel_approx}). Thus, we have shown that the same phase response for light propagating over a distance $d_{\mathrm{eff}}$ in free space can instead be achieved using a photonic crystal slab, provided that the band structure of the slab is isotropic as described by Eq.~(\ref{eq:dispersion}) with the parameters that satisfy 
\begin{equation}
    \label{eq:prop_dist}
    d_{\mathrm{eff}} = \frac{4\pi \alpha}{\lambda\gamma_0 (1+\Omega_0^2)}.
\end{equation}
With a choice of small $\gamma_0$, (i.e.~using guided resonances with high quality factors), one can achieve a large $d_{\mathrm{eff}}$ using a photonic crystal slab with a physical thickness that is far smaller than $d_{\mathrm{eff}}$. Therefore, we have shown that the use of a properly designed photonic crystal slab can achieve the squeezing of free space.  

As was noted in \cite{Guo2018}, in a typical photonic crystal slab, due to the photonic spin-orbit coupling, the band structure near $\bm{k} = \bm{0}$ is not isotropic. However, one can use the specific design procedure as discussed in \cite{Guo2018} to design a photonic crystal slab with an isotropic band structure of the form of Eq.~(\ref{eq:dispersion}). In the next section, we will follow the same design procedure of Ref.~\cite{Guo2018} in our numerical design. 

\section{\label{sec:numerical}Numerical demonstration}
Based on the theoretical considerations above, we provide a concrete design of such a photonic crystal slab device. Our device consists of three layers as shown in Fig.~\ref{fig:scheme}c. The middle layer is a photonic crystal slab with a lattice constant $a$. It has a thickness of $d=0.55a$, and contains a square array of circular holes with radius $r=0.111a$. Two homogeneous slabs with thickness $d_s=0.07a$ are placed symmetrically besides  the photonic crystal slab. The air gaps between the middle slab and the homogeneous slabs are $d_g=0.94a$. The total thickness of such a device is $d_{T} = d+2d_s+2d_g = 2.57a$. All slabs are made of materials with a permittivity $\varepsilon=12$, which approximates that of Si or GaAs in the infrared wavelength range. 

\begin{figure}[htb]
    \centering
    \includegraphics[width=0.98\columnwidth]{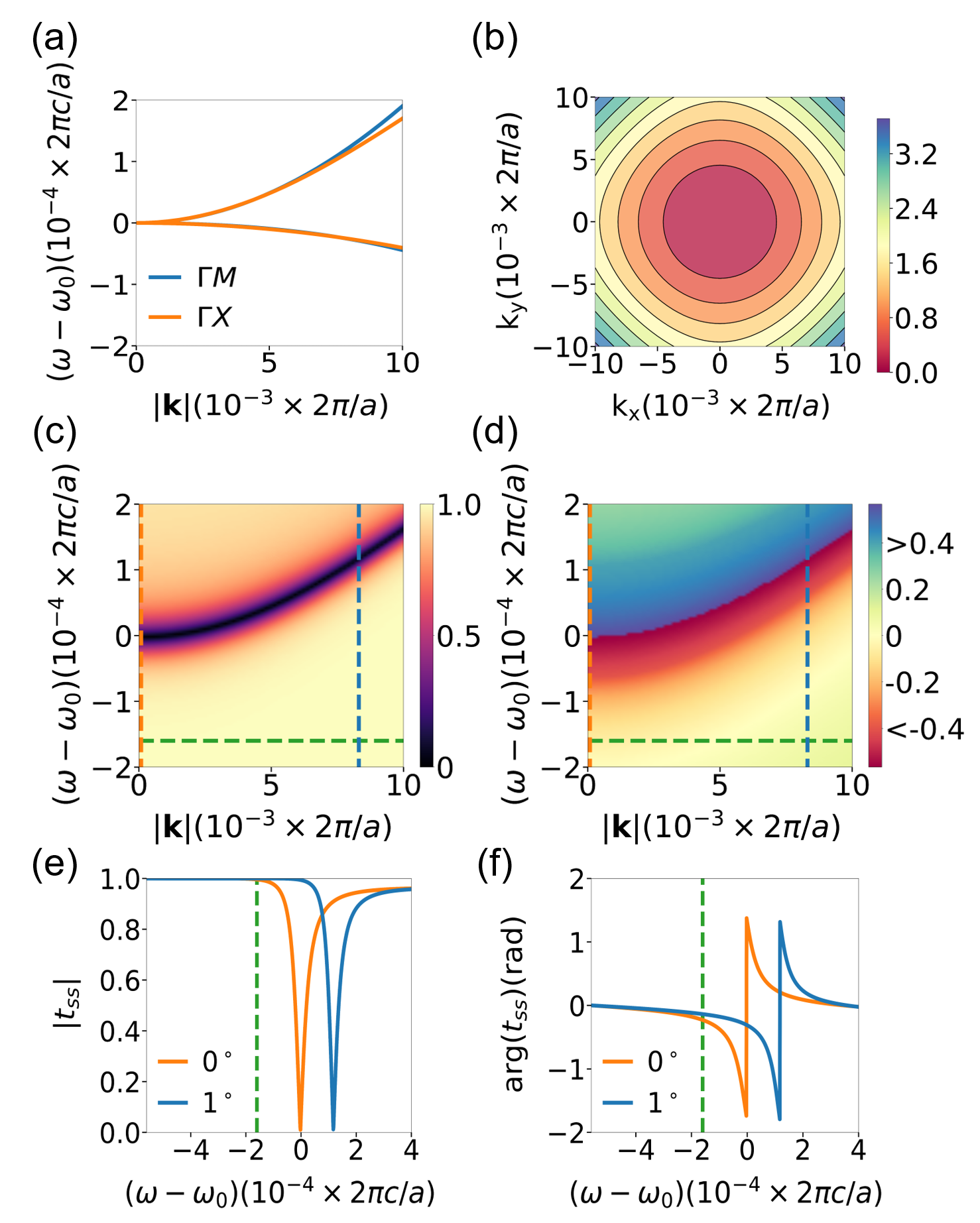}
    \caption{The band structure and the transmission properties of the photonic crystal slab device shown in Fig.~\ref{fig:scheme}c. (a) Band structure of guided resonances that couples to near normal incident light, shown along $\Gamma X$ and $\Gamma M$ directions. (b) Isofrequency contours of the upper band. (c) Transmission magnitude $|t_{ss}|$ of the incident $s$ polarized light at different frequencies and in-plane wavevectors. (d) Transmission phase $\operatorname{arg}(t_{ss})$ of the incident $s$-polarized light at different frequencies and in-plane wavevectors. (e) $|t_{ss}|$ as a function of frequency, for $s$-polarized light with an incident angle $0^\circ$ and $1^\circ$. (f) $\operatorname{arg}(t_{ss})$ as a function of frequency, for $s$ polarized light with an incident angle $0^\circ$ and $1^\circ$. In (c-f), the orange and blue lines correspond to the $0^\circ$ and $1^\circ$ incidence, respectively, and the green line indicates the operating frequency $\omega=0.47640\times 2\pi c/a$.}
    \label{fig:band}
\end{figure}
Such a photonic crystal slab device hosts a pair of guided resonances that are doubly degenerate at the $\Gamma$ point ($\bm{k}=\bm{0}$) with the frequency $\omega_0 = 0.47656 \times 2\pi c/a$. (Such a two-fold degeneracy is required in order for the guided resonance to couple to normally incident light~\cite{Fan2002}.) As we mentioned above, in general, the band structure of the guided resonances is anisotropic around $\bm{k}=\bm{0}$ for a photonic crystal slab with the $C_{4v}$ symmetry. However, with the geometry parameters chosen above, which are obtained following the same procedure as in Ref.~\cite{Guo2018}, we show the corresponding band structure as obtained using the guided mode expansion method~\cite{andreani2006, minkov2014} in Fig.~\ref{fig:band}. Fig.~\ref{fig:band}a shows that for each band, the dispersion along the $\Gamma X$ coincides with that along the $\Gamma M$ direction. Fig.~\ref{fig:band}b shows that the isofrequency contours for the upper band are almost circular. Both bands become almost completely isotropic:  
\begin{equation}
    \label{eq:w_k_2bands}
    \omega_i(\bm{k}) \approx \omega_0 + \alpha_i (k_x^2+k_y^2), \ i=1,2, 
\end{equation}
where $\alpha_1=1.79\,ca/(2\pi)$, $\alpha_2=-0.43\,ca/(2\pi)$ and the $1,2$ subscripts correspond to the upper and lower band, respectively (cf.~Eq.~(\ref{eq:dispersion})). Also, we note that the dispersion of radiative linewidths $\gamma_i(\bm{k})$ is anisotropic. Nonetheless,  the dispersion of $\gamma_i(\bm{k})$ is  much smaller than that of  $\omega_i(\bm{k})$ so that its effect on transfer function is negligible: $\gamma_i(\bm{k})\approx \gamma_0$ (cf.~Eq.~(\ref{eq:gamma})).

As was noted in Ref.~\cite{Guo2018}, the isotropic band structure leads to the remarkable effect of \emph{single-band excitation}: $s$/$p$-polarized light only couples to the upper/lower band, respectively, for every direction of incidence~\cite{Guo2018}. Consequently, the $s$ or $p$ polarization is preserved upon transmission  through the device ($t_{ps}(\omega,\bm{k})=0$). We calculate the transmittance of $s$-polarized light $t_{ss}(\omega,\bm{k})$ by the Fourier Modal Method using a freely available software package~\cite{liu2012c}. Figs.~\ref{fig:band}c and \ref{fig:band}d depict the magnitude  $|t_{ss}|$ and phase $\operatorname{arg}(t_{ss})$, respectively, at a general azimuthal angle $\phi = \arctan{(k_y/k_x)}=14^\circ$. Due to the isotropic band structure, the results are essentially the same for any other $\phi$. The plots clearly show that $s$-polarized light only excites the upper band. The transmission exhibits sharp dip in magnitude and rapid variation in phase near the band dispersion of the guided resonances. 

Figs.~\ref{fig:band}e and \ref{fig:band}f plot the spectra of the transmission magnitude and phase at incident angles $\theta=0^\circ$ and $\theta=1^\circ$, respectively. For both incident angles, both the  amplitude and phase follows the Fano lineshape formula as described by Eq.~(\ref{eq:Fano_Omega}) and Eq.~(\ref{eq:t_arg}). As $\theta$ increases, the resonance shifts to higher frequencies, in consistency with the band structure. Based on Figs.~\ref{fig:band}e and \ref{fig:band}f, we choose the  operating frequency at $\omega_{\mathrm{op}}=0.47640\times 2\pi c/a$ as indicated by the green dashed line. At this frequency, the transmission coefficient magnitude stays close to unity while the phase varies  substantially for varying  incident angles, as is required for Eq.~(\ref{eq:t_kx_ky}). Also, we note that at this frequency the corresponding wavelength is greater than the lattice constant such that there is no diffraction due to the slab.

\begin{figure}[htbp]
    \centering
    \includegraphics[width=0.98\columnwidth]{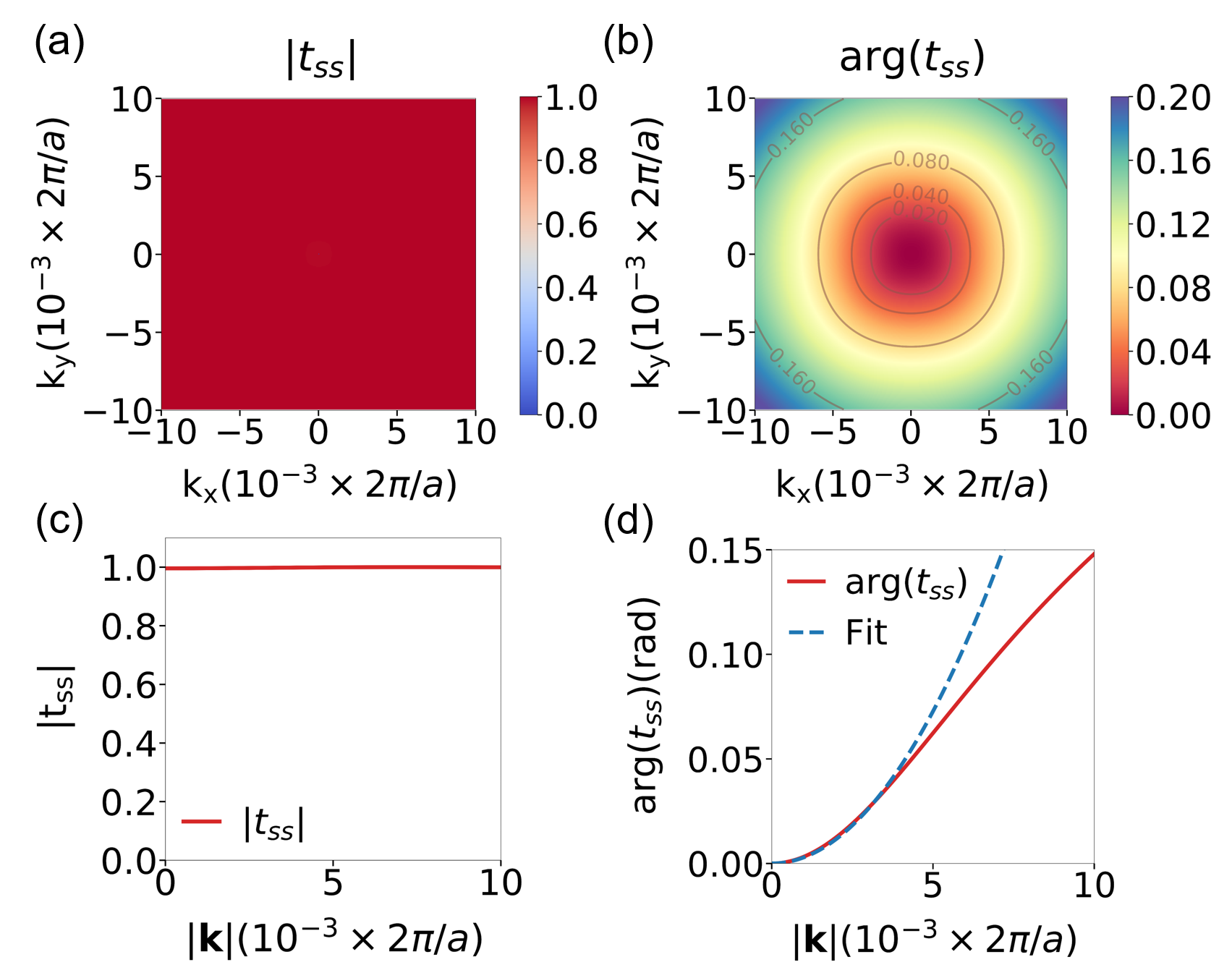}
    \caption{Transfer function of the photonic crystal slab device. (a) $|t_{ss}|$ at different in-plane wavevectors $(k_x,k_y)$. (b) $\operatorname{arg}(t_{ss})$ at different in-plane wavevectors $(k_x,k_y)$. (c) $|t_{ss}|$ as a function of in-plane wavevector magnitude $|\bm{k}|$. (d) $\operatorname{arg}(t_{ss})$ as a function of in-plane wavevector magnitude $|\bm{k}|$, along with a fit to quadratic function.}
    \label{fig:transfer}
\end{figure}

At the operating frequency $\omega_{\mathrm{op}}$, Figs.~\ref{fig:transfer}a and b show the magnitude and phase of the transmission $t_{ss}(k_x, k_y)$ in the $k_x$ and $k_y$ plane. Figs.~\ref{fig:transfer}c and d show the magnitude and phase of the transmission along the $\Gamma X$ direction.  The transmission has a magnitude of unity over the entire wavevetor range considered (Fig.~\ref{fig:transfer}a). Hence the structure behaves as an all-pass filter in this wavevector range. The phase is isotropic and shows a quadratic dependency of $|\bm{k}|$. The magnitude and phase response therefore agrees with Eq.~(\ref{eq:t_kx_ky}).

By fitting the phase response with a quadratic function in $|\bm{k}|$, as shown in Fig. 3d, we obtain the parameter $\alpha/[\gamma_0 (1+\Omega_0^2)] = 61.97\,a^2$ for Eq.~(\ref{eq:t_kx_ky}), thus the device is equivalent to free space of a thickness   $d_{\mathrm{eff}}=371\,a$ as determined from Eq.~(\ref{eq:prop_dist}). Since the device has a thickness of $d_T = 2.57\,a$, by replacing the free space with this device we have preserved the angle-dependent phase response of the free space while reducing the required physical thickness by a compression ratio of $d_{\mathrm{eff}}/d_T = 144$. Moreover, since the device behaves as an all-pass filter, a longer effective propagation distance can be achieved by simply cascading multiple devices together. A cascade of $N$ devices results in an effective propagating distance of $Nd_{\mathrm{eff}}$, whereas the compression ratio is unchanged from a single device.


\begin{figure}[htbp]
    \centering
    \includegraphics[width=0.94\columnwidth]{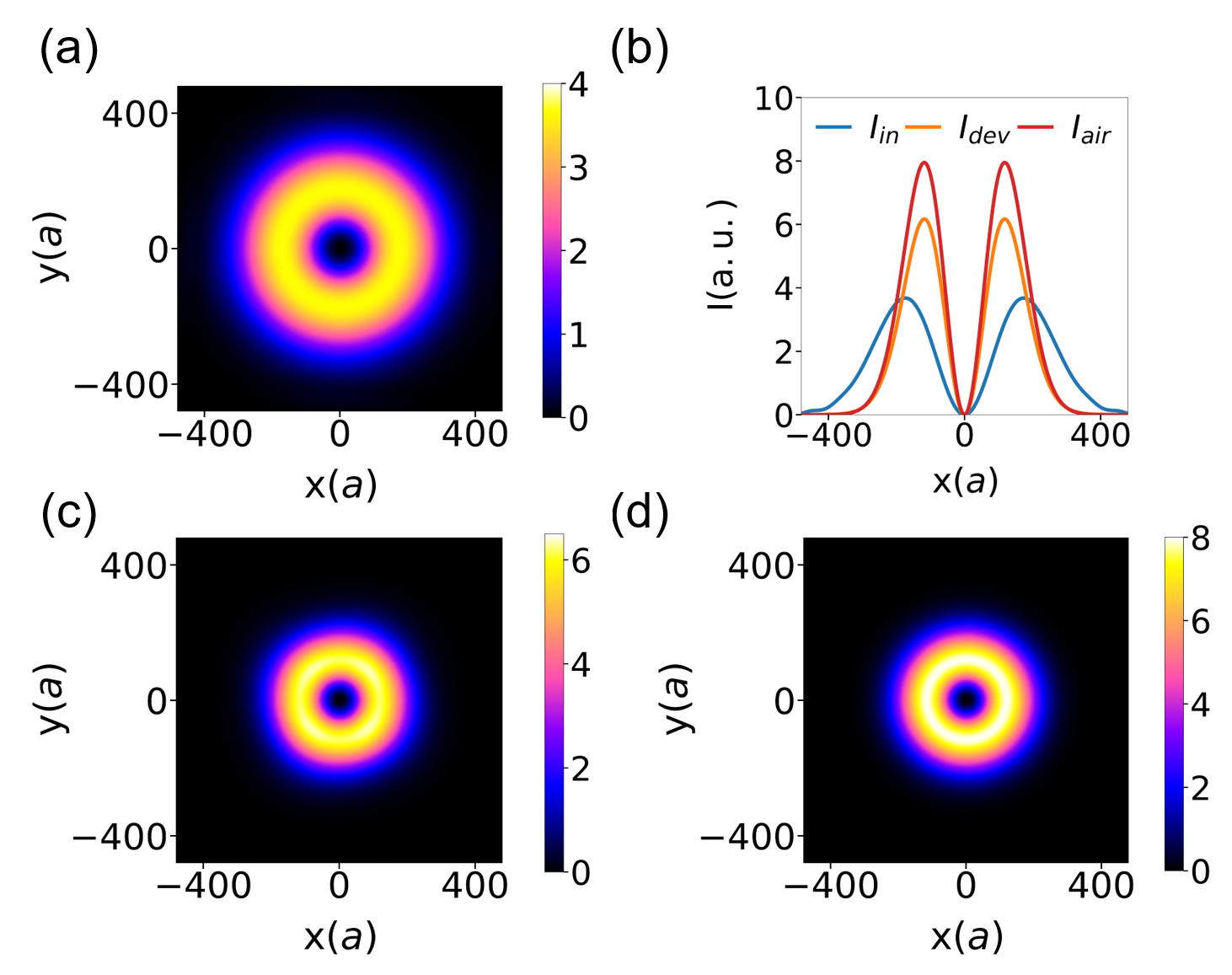}
    \caption{Demonstration of the device with actual beam propagation. (a) The intensity distribution of an input converging Laguerre-Gaussian beam with $l=1,m=0$. (b) The radial intensity profile of the three beams in (a), (c), and (d). (c) The intensity distribution of the input beam immediately after passing through $40$ devices with a total thickness $103a$. (d) The intensity distribution of the input beam after a free space propagation of $14823a$.}
    \label{fig:demo}
\end{figure}

In Fig.~\ref{fig:demo}, we provide a numerical demonstration of  the performance of our device \cite{J.S.Schmidt2014}. We consider a converging  azimuthally ($s$) polarized Laguerre-Gaussian beam with $l=1,m=0$ at the operating frequency $\omega_{\mathrm{op}}$. Its intensity distribution at $z=0$ plane is shown in Fig.~\ref{fig:demo}a. For demonstration, we directly simulate a cascade of $N=40$ devices with an air gap of thickness $0.5 a$ between every two neighboring devices. The total thickness of the $40$ cascaded devices including the air gaps are $123\,a$.
We numerically  calculate the intensity distribution of the transmitted beam after the cascaded devices as shown in Fig.~\ref{fig:demo}c.  The beam size significantly reduces after passing through the devices. We also numerically verify that the result for $40$ devices is indeed the same as that for a single device repeated $40$ times. 
As comparison, in Fig.~\ref{fig:demo}d we plot the intensity distribution of the original beam after the propagation in free space by $40\,d_{\mathrm{eff}}=14823\,a$. Fig.~\ref{fig:demo}c and Fig.~\ref{fig:demo}d agree very well. In Fig.~\ref{fig:demo}b we plot the radial profile for the input beam and the aforementioned two output beams. We see that our device can substitute free space propagation and reproduce the beam width and beam shape very well. The reduced peak intensity is caused by the deviation of the transfer function at large wavevectors (Fig.~\ref{fig:transfer}d).


\section{\label{sec:conclusion}Discussion and Conclusion}
We note that our device, like free space, is invariant under both translation along the longitudinal direction (i.e. the propagation direction or $z$-direction in Fig.~\ref{fig:scheme}), and transverse (i.e.~within the $x$-$y$ plane) directions. Therefore the device can be inserted anywhere along the optical path and the effect is independent of the location of insertion. Our device works for both spatially coherent and incoherent optical waves. We also note that our device can work at different frequencies as well, though the equivalent propagation distance can be different. Correction of such a chromatic aberration is of interest for future research.

In conclusion, we have shown that free space can be substituted with \emph{nonlocal} flat optics. We derive the general criteria for replacing free space with an optical device, and provide a concrete design of a photonic crystal slab device. Such a device can substitute much thicker free space. Our work provides an important complement to local flat optics and may prove important for miniaturization of optical systems.

\begin{acknowledgments}
We wish to acknowledge Dr.~Meng Xiao for helpful discussion. This work is supported by an U.~S.~Air Force Office of Scientific Research (AFOSR) MURI grant (Grant No. FA9550-17-1-0002), and by a Vannevar Bush Faculty Fellowship from the U.~S.~Department of Defense. (N00014-17-1-3030). While we were finalizing this manuscript, a similar recent preprint was brought to our attention
(Ref.~\cite{reshef2020}). Our design here may result in a more compact design as compared with the multilayer configuration shown in Ref.~\cite{reshef2020}.

\end{acknowledgments}

\bibliography{References}

\end{document}